\documentclass[]{rsos}%%%%where rsos is the template name

\usepackage{tabularx,ragged2e,booktabs,caption}
\newcommand{\ra}[1]{\renewcommand{\arraystretch}{#1}}

\begin{document}

%%%% Article title to be placed here
\title{Boundary regularised integral equation formulation of the Helmholtz equation in acoustics}

\author{%%%% Author details
Qiang Sun$^{1}$, Evert Klaseboer$^{2}$, Boo-Cheong Khoo$^{1}$ \newline and Derek Y. C. Chan$^{1, 2, 3, 4}$}

%%%%%%%%% Insert author address here
\address{$^{1}$Department of Mechanical Engineering, National University of Singapore, 10 Kent Ridge Crescent, 119260, Singapore \\
$^{2}$Institute of High Performance Computing, 1 Fusionopolis Way, 138632, Singapore\\
$^{3}$Department of Mathematics and Statistics, The University of Melbourne, Parkville 3010 VIC Australia\\
$^{4}$Department of Chemistry and Biotechnology, Swinburne University of Technology, Hawthorn 3122 VIC Australia}

%%%% Subject entries to be placed here %%%%
\subject{applied mathematics, boundary integral method, acoustics}

%%%% Keyword entries to be placed here %%%%
\keywords{boundary integral equation, Helmholtz equation, singularity removal, wave equation}

%%%% Insert corresponding author and its email address}
\corres{Derek Y. C. Chan\\
\email{D.Chan@unimelb.edu.au}}

%%%% Abstract text to be placed here %%%%%%%%%%%%
\begin{abstract}
A boundary integral formulation for the solution of the Helmholtz equation is developed in which all traditional singular behaviour in the boundary integrals is removed analytically. The numerical precision of this approach is illustrated with calculation of the pressure field due to radiating bodies in acoustic wave problems. This method facilitates the use of higher order surface elements to represent boundaries, resulting in a significant reduction in the problem size with improved precision. Problems with extreme geometric aspect ratios can also be handled without diminished precision. When combined with the CHIEF method, uniqueness of the solution of the exterior acoustic problem is assured without the need to solve hypersingular integrals.
\end{abstract}
%%%%%%%%%%%%%%%%%%%%%%%%%%%

%%%%%%%%%% Insert the texts which can accomdate on firstpage in the tag "fmtext" %%%%%

\begin{fmtext}

%=============================================================

\section{Introduction}
Central to acoustic wave theory is solving the Helmholtz equation for the pressure field in the frequency domain. The boundary integral method is commonly used because of the reduction in spatial dimension. However, it is well-known that the numerical solution of the \linebreak

\end{fmtext}

%%%%%%%%%%%%%%% End of first page %%%%%%%%%%%%%%%%%%%%%

\maketitle

{\setlength{\parindent}{0ex}
boundary integral equation for external problems can become inaccurate when the wave number is close to one of the eigenvalues of the internal problem \cite{Helmholtz1896, Rayleigh1896}. The issue has been addressed by two common methods. The CHIEF method due to Schenck \cite{Schenck1968} imposes an additional constraint on the solution of the boundary integral equation by requiring it to vanish at selected positions inside the boundary to suppress the resonant solution that has no physical significance in the external problem. An alternative approach proposed by Burton and Miller \cite{Burton1971} involved solving a hypersingular integral equation in which the original boundary integral equation is the real part and the boundary integral equation for the normal derivative is the imaginary part. In both cases, the equations to be solved contain mathematical singularities associated with the conventional formulation of the boundary integral equation and much effort since then has been concerned with the expeditious and efficient treatment of these singularities \cite{Gaul2003}.}

Recently we reformulated the boundary integral solution of the Laplace equation: $\nabla^2\phi = 0$, and the Stokes equation of fluid mechanics whereby all singular terms in the integrals are removed analytically \cite{Klaseboer2012}. This regularisation of all the singular behaviour on the boundary means that the surface integrals can be evaluated using any convenient quadrature method. A significant practical consequence of this is that high numerical precision can be obtained with mixed boundary conditions \cite{Sun2013} and for problems with multiscale characteristics such as those with boundaries that are very close together compared to their characteristic dimensions or where the boundaries possess extreme geometric aspect ratios \cite{Sun2014}.

In this work, we apply this boundary regularisation to the Helmholtz equation that removes much of the technical effort needed to use linear or quadratic surface elements to represent boundaries and results in significant improvement in the accuracy in the evaluation of surface integrals. Also, it is no longer necessary to calculate the solid angle at each node -- a complexity that can discourage the use of higher order surface elements. As a result, far fewer degrees of freedom are needed to achieve the same precision that in turn translates to a significant decrease in computational time -- demonstrated here by numerical examples. When the high precision of this boundary regularised formulation is combined with the CHIEF method \cite{Schenck1968}, the possibility of numerical error arising from the resonance solution becomes negligible in practice. The framework given here is applicable for both external and internal problems, for Dirichlet, Neumann or mixed boundary conditions and for radiation as well as scattering problems.

Although examples motivated by problems in acoustics have been used to demonstrate the theoretical formulation and practical numerical advantages, this method of de-singularising boundary integral equations can be applied in other contexts such as the Laplace equation \cite{Sun2014}, the equations of hydrodynamics \cite{Klaseboer2012, Sun2013} and linear elasticity or any equation that belongs to the Moisil-Theodorescu system \cite{Zabarankin2012}.

%=============================================================

\section{Boundary regularised integral equation formulation (BRIEF)}
To illustrate the boundary regularised integral equation formulation of the Helmholtz equation, we draw on the problem of calculating the acoustic pressure wave generated by a vibrating boundary specified by a closed surface, or more generally a set of closed surfaces denoted by $S$. The acoustic pressure, $p(\mathbf{x})$ outside $S$, obeys the Helmholtz equation: $\nabla ^ 2 p + k ^ 2 p = 0$ where $k$ is the wave number. The solution can be found by solving the conventional boundary integral equation that follows from using Green's second identity~\cite{Brebbia1992}
\begin{align}\label{eq:cnbimhlmhltz}
c_0 \, p(\mathbf{x}_0) + \int_{S + S_{\infty}} p(\mathbf{x}) \frac{\partial G(\mathbf{x}_0, \mathbf{x})}{\partial n} \text{d}S(\mathbf{x})  = \int_{S+ S_{\infty}}  \frac{\partial p(\mathbf{x})}{\partial n} G(\mathbf{x}_0, \mathbf{x}) \text{d}S(\mathbf{x}),
\end{align}
where $G(\mathbf{x}_0, \mathbf{x})=e^{ikr}/r$, with $r \equiv |\mathbf {x} - \mathbf{x}_{0}|$, is the fundamental solution of the Helmholtz equation. These integrals are taken over the set of closed surfaces, $S$, specified by the problem and also over the ``surface at infinity'', $S_{\infty}$ at which the Sommerfeld radiation condition is applied (see the Appendix). The points $\mathbf{x}$ and $\mathbf{x}_0$ are on the boundaries. For simplicity, we assume that the surfaces $S$ and $S_{\infty}$ have well-defined tangent planes at all points on the surfaces. We refer the readers to our another work \cite{Sun2014} for the treatment of surfaces with sharp edges and vertices. The normal derivatives are defined by $\partial p / \partial n \equiv \nabla p \cdot \mathbf{n}(\mathbf{x})$ and $\partial G / \partial n \equiv \nabla G \cdot \mathbf{n}(\mathbf{x})$ where $\mathbf{n}(\mathbf{x})$ is the unit normal vector at $\mathbf{x}$ pointing out of the solution domain. The solid angle, $c_0$ at $\mathbf{x}_{0}$ is equal to $2 \pi$ if the tangent of the boundary at $\mathbf{x}_{0}$ is defined, otherwise it has to be calculated from the local geometry \cite{Mantic93}. Equation (\ref{eq:cnbimhlmhltz}) provides a relation between the function $p$ and its normal derivative $\partial p / \partial n$ on the surfaces $S$ and $S_\infty$. For Dirichlet (Neumann) problems, $p$ ($\partial p / \partial n$) is specified on the surfaces, and (\ref{eq:cnbimhlmhltz}) can be solved for $\partial p / \partial n$ ($p$).

In spite of the fact that the physical problem may be well-behaved on the boundaries, the mathematical singularities in (\ref{eq:cnbimhlmhltz}) at $\mathbf{x} = \mathbf{x}_{0}$ due to $G$ and $\partial G/\partial n$ require careful treatment. To quote Jaswon \cite{Jaswon1963} and Symm \cite{Symm1963}: ``Most integral equations of physical significance involve singular, or weakly singular, kernels, thereby hampering the procedures of both theoretical and numerical analysis." -- a statement that is still valid up to this day.

Here we seek to eliminate these singularities analytically by exploiting the linear nature of (\ref{eq:cnbimhlmhltz}) as follows.

But before we tackle (\ref{eq:cnbimhlmhltz}) that corresponds to the Helmholtz equation for $p(\mathbf{x})$, consider first the simpler case of the Laplace equation: $\nabla ^ 2 q(\mathbf{x}) = 0$ for which the corresponding boundary integral equation is
\begin{align}\label{eq:cnbimlaplace}
c_0 \, q(\mathbf{x}_0) + \int_{S + S_{\infty}} q(\mathbf{x}) \frac{\partial G_0(\mathbf{x}_0, \mathbf{x})}{\partial n} \text{d}S(\mathbf{x})  = \int_{S+ S_{\infty}}  \frac{\partial q(\mathbf{x})}{\partial n} G_0(\mathbf{x}_0, \mathbf{x}) \text{d}S(\mathbf{x}),
\end{align}
where $G_0(\mathbf{x}_0, \mathbf{x})=1/|\mathbf {x} - \mathbf{x}_{0}|$. The standard way to remove the singularity arising from $\partial G_0/\partial n$ on the left hand side is to note that the constant $q(\mathbf{x}_0)$ also satisfies the Laplace equation with a vanishing normal derivative on the boundaries, so that equation (\ref{eq:cnbimlaplace}) for $[q(\mathbf{x}) - q(\mathbf{x}_0)]$ becomes
\begin{align}\label{eq:cnbimlaplace2}
 \int_{S + S_{\infty}} [q(\mathbf{x}) - q(\mathbf{x}_0)] \frac{\partial G_0(\mathbf{x}_0, \mathbf{x})}{\partial n} \text{d}S(\mathbf{x})  = \int_{S+ S_{\infty}}  \frac{\partial q(\mathbf{x})}{\partial n} G_0(\mathbf{x}_0, \mathbf{x}) \text{d}S(\mathbf{x}).
\end{align}
This is known as the 'constant value subtraction'. However, the integral on the right hand side of (\ref{eq:cnbimlaplace2}) still involves an integration over the singularity from $G_0$ and is normally handled by a change to local polar coordinates \cite{Gaul2003}. However, this singularity can also be removed analytically \cite{Sun2014}. Implicit in this approach is the generally valid assumption that as $\mathbf{x} \rightarrow \mathbf{x}_0$, $[q(\mathbf{x}) - q(\mathbf{x}_0)]$ vanishes as $|\mathbf{x} - \mathbf{x}_0|$ or faster.

In this paper, we extend the approach \cite{Sun2014} developed for the Laplace equation to remove \emph{all} singularities in the boundary integral equation (\ref{eq:cnbimhlmhltz}) for the Helmholtz problem. However, since a constant is not a solution of the Helmholtz equation, we need to construct an auxiliary function $\psi(\mathbf {x})$
for a given value of $\mathbf{x}_0$ in (\ref{eq:cnbimhlmhltz}) that satisfies the Helmholtz equation that is analogous to the constant, $q(\mathbf{x}_0)$, in the case of the Laplace equation. We choose $\psi(\mathbf {x})$ to be linear in the pressure, $p(\mathbf{x}_0)$ and its normal derivative, $(\partial p / \partial n)_0$ at $\mathbf{x}_0$
\begin{align}\label{eq:hlmhltz}
\psi(\mathbf {x}) \equiv p(\mathbf{x}_0) \, g({\mathbf{x}}) + \left(\frac{\partial p}{\partial n}\right)_{0}f({\mathbf{x}}).
\end{align}
Now provided the general functions $g(\mathbf{x})$ and $f(\mathbf{x})$ satisfy the Helmholtz equation with the following boundary conditions:
\begin{subequations}
\label{eq:nsbimcndtn}
\begin{align}
  \label{eq:nsbimcndtn_f1}
  &\nabla^2 g(\mathbf{x}) + k^2 g(\mathbf{x})= 0, \qquad g(\mathbf{x}_0) = 1, \qquad \nabla g(\mathbf{x}_0) \cdot \mathbf{n}_0 = 0, \\
  \label{eq:nsbimcndtn_f2}
  &\nabla^2 f(\mathbf{x}) + k^2 f(\mathbf{x})= 0, \qquad f(\mathbf{x}_0) = 0, \qquad \nabla f(\mathbf{x}_0) \cdot \mathbf{n}_0 = 1,
\end{align}
\end{subequations}
with $\mathbf{n}_0 \equiv \mathbf{n}(\mathbf{x}_0)$ being the outward unit normal at $\mathbf{x}_0$, $\psi(\mathbf {x})$ will also satisfy the same boundary integral equation as (\ref{eq:cnbimhlmhltz}). By taking the difference between the boundary integral equations for $p(\mathbf{x})$ and $\psi(\mathbf{x})$, we obtain
\begin{align} \label{eq:BRIEF}
&\int_{S + S_{\infty}} \left[p(\mathbf{x}) - p(\mathbf{x}_0)g({\mathbf{x}})  -  \left(\frac{\partial p}{\partial n}\right)_{0} f({\mathbf{x}}) \right] \frac{\partial G(\mathbf{x}_0, \mathbf{x})}{\partial n} \text{d}S(\mathbf{x})  \nonumber \\
&= \int_{S + S_{\infty}} \left[ \frac{\partial p(\mathbf{x})}{\partial n} - p(\mathbf{x}_0) \nabla g(\mathbf{x}) \cdot \mathbf{n}(\mathbf{x}) - \left(\frac{\partial p}{\partial n}\right)_{0} \nabla f(\mathbf{x}) \cdot \mathbf{n}(\mathbf{x}) \right] G(\mathbf{x}_0, \mathbf{x}) \text{d}S(\mathbf{x}).
\end{align}
The conditions imposed on $g(\mathbf{x})$ and $f(\mathbf{x})$ at $\mathbf{x}_0$ in (\ref{eq:nsbimcndtn}) will remove \emph{both} the singularities due to $G$ and $\partial G / \partial n$ in the new boundary integral equation in (\ref{eq:BRIEF}) under the generally valid assumption that as $\mathbf{x} \rightarrow \mathbf{x}_0$, $\left[ p(\mathbf{x}) - \psi (\mathbf{x}) \right]$ and $\left[ \partial p /\partial n - \partial \psi /\partial n \right]$ vanish as $|\mathbf{x} - \mathbf{x}_0|$ or faster. Therefore, if $g(\mathbf{x})$ and $f(\mathbf{x})$ satisfy (\ref{eq:nsbimcndtn}), all singular behaviour in (\ref{eq:BRIEF}) due to $G$ and $\partial G/ \partial n$ will be removed. The formal proof is a straightforward generalisation of that given in \cite{Klaseboer2012} for the Laplace equation.

Equation (\ref{eq:BRIEF}) is the final key result for the singularity-free~\cite{Klaseboer2012} boundary regularised integral equation formulation (BRIEF) of the solution of the Helmholtz equation that replaces the conventional boundary integral equation in (\ref{eq:cnbimhlmhltz}). The integrals are now completely free of singularities \cite{Klaseboer2012} and can be evaluated by any convenient integration quadrature. Given values for $p$ (Dirichlet) or $\partial p / \partial n$ (Neumann) or a relation between these two quantities on the boundary, $S$, (\ref{eq:BRIEF}) can be readily solved. Also the term involving the solid angle, $c_0$, in the conventional boundary integral equation, (\ref{eq:cnbimhlmhltz}), has now been eliminated. Thus higher order area elements -- linear, quadratic or splines, can be used to represent the surface to improve numerical accuracy without the need to calculate the solid angle at each node.

For our key result in (\ref{eq:BRIEF}), we see that apart from having to satisfy (\ref{eq:nsbimcndtn}), there is considerable freedom in selecting the functional forms of $g(\mathbf{x})$ and $f(\mathbf{x})$ and hence $\psi(\mathbf{x})$. As a specific example, we can choose the following for $g(\mathbf{x})$ and $f(\mathbf{x})$
\begin{align}  \label{eq:gANDf}
g({\mathbf{x}}) = \frac{a\cos[k(r_d-a)]}{r_{d}} + \frac{\sin[k(r_d-a)]}{k\,r_{d}}, \quad  f({\mathbf{x}}) = \frac{a \sin[k(r_d-a)]}{b\, k\,r_{d}},
\end{align}
where $\mathbf{x}_d$ is any convenient point \emph{outside} the solution domain in order to ensure $\psi(\mathbf{x})$ is non-singular within the solution domain, with $a \equiv |\mathbf{x}_{0} - \mathbf{x}_{d}|$, $r_d \equiv |\mathbf{x} - \mathbf{x}_{d}|$ and $b \equiv (\mathbf{x}_{0} - \mathbf{x}_{d})\cdot \mathbf{n}(\mathbf{x}_0)/|\mathbf{x}_{0} - \mathbf{x}_{d}| \neq 0$ (see figure 1a). For this particular choice of $g(\mathbf{x})$ and $f(\mathbf{x})$, the integral over the surface at infinity, $S_{\infty}$ in (\ref{eq:BRIEF}) can be found analytically using the Sommerfeld radiation condition~\cite{Sommerfeld12, Schot92}, (see Appendix \ref{sec:appA} for details)
\begin{eqnarray}
\label{eq:intinf1}
p(\mathbf{x}_0)\int_{S_{\infty}} \left[\nabla g(\mathbf{x}) \cdot \mathbf{n}\, G - g({\mathbf{x}}) \frac{\partial G}{\partial n} \right]  \text{d}S &=& 2 \pi p(\mathbf{x}_0)\left( 1 + \frac{i} {ka}\right) \left[ 1 - e^{2ika} \right], \\
%\end{align}
%\begin{align}
\label{eq:intinf2}
\left(\frac{\partial p}{\partial n}\right)_{0}\int_{S_{\infty}}  \left[\nabla f(\mathbf{x}) \cdot \mathbf{n}\, G - f({\mathbf{x}}) \frac{\partial G}{\partial n} \right]  \text{d}S &=& \left(\frac{\partial p}{\partial n}\right)_{0} \frac {2 \pi i}{kb}\left[ 1 - e^{2ika} \right].
\end{eqnarray}

In the limit $k = 0$, the Helmholtz equation reduces to the Laplace equation with $g(\mathbf{x}) \rightarrow 1$ and $f(\mathbf{x}) \rightarrow (a/b) [1-a/|\mathbf{x} - \mathbf{x}_d|] $, and (\ref{eq:BRIEF}) will become a boundary regularised integral equation for the solution of the Laplace equation given earlier \cite{Sun2014}. This is an example of how the singularity in the integral on the right hand side of (\ref{eq:cnbimlaplace2}) may be removed analytically.

We note that the conventional boundary integral formulation of the solution of the Helmholtz equation, or for that matter solutions of the Laplace equation or the equations of hydrodynamics or elasticity, gives rise to singularities in the integrands of the surface integrals. These mathematical singularities originate from the fundamental solution of the equations used in Green's second identity whereas the physical problem is perfectly well-behaved on the boundaries. Therefore it is not too surprising that these mathematical singularities can be completely removed by subtracting a related auxiliary solution of the governing equation: $\psi(\mathbf{x})$ in the case of the Helmholtz equation considered here. Apart from having to satisfy the governing differential equation and some mild constraints, see (\ref{eq:nsbimcndtn}), there is considerable flexibility in choosing the precise functional form of the auxiliary solution or any free parameters that it may contain - such as the value of $\mathbf{x}_d$ in (\ref{eq:gANDf}). A broad physical interpretation is that we have constructed a pressure field, $\psi(\mathbf{x})$, that cancels the value of $p$ and $\partial p/\partial n$ at $\mathbf{x}_0$.

One well-studied uniqueness issue is that the solution of the boundary integral method, (\ref{eq:cnbimhlmhltz}), with a closed boundary is identical to that obtained from solving directly the differential form of the Helmholtz equation: $\nabla ^ 2 p + k ^ 2 p = 0$, except at a discrete set of values of $k$. These $k$ values correspond to the resonant wave numbers of the closed boundary \cite{Schenck1968, Burton1971}. The solution of the BRIEF, (\ref{eq:BRIEF}), also shares this feature. The resonant or homogeneous solutions corresponding to the resonant values (eigenvalues) of the closed boundary will always emerge from solving the integral equation. But as we shall see in Section 5, the higher precision that the BRIEF can attain will alleviate much of the practical numerical difficulty.

% ======================================================

\section{Accurate evaluation at field points near boundaries}
The boundary regularised integral equation formulation (BRIEF) of the solution of the Helmholtz equation also offers an accurate and numerically robust method to calculate the solution at field points close to boundaries. Indeed, the loss of precision due to the near singular behaviour in the evaluation of the requisite integrals that arise in the conventional boundary integral method is often more difficult to deal with than the singularities on the boundaries.

To evaluate the solution $p(\mathbf{x}_p)$ of the Helmholtz equation at a point $\mathbf{x}_p$ in the solution domain, we first use the conventional boundary integral equation to find $[p(\mathbf{x}) - \psi(\mathbf{x})]$ at $\mathbf{x}_p$, with $\mathbf{x}_0$ in (\ref{eq:hlmhltz}) taken to be the node on the boundary closest to $\mathbf{x}_p$
\begin{align} \label{eq:CBIMpminuspsi}
4\pi p(\mathbf{x}_p) =    4\pi \psi(\mathbf{x}_p)   - \int_{S + S_{\infty}}  & \left[ p(\mathbf{x}) -  \psi(\mathbf{x}) \right]  \frac{\partial G(\mathbf{x}_p, \mathbf{x})}{\partial n} \text{d}S(\mathbf{x})  \nonumber \\
& + \int_{S + S_{\infty}}  \frac{\partial [p(\mathbf{x}) - \psi(\mathbf{x})]} {\partial n} G(\mathbf{x}_p, \mathbf{x}) \text{d}S(\mathbf{x}).
\end{align}
The near singular behaviour of the surface integrals due to $G$ and $\partial G / \partial n$ as $\mathbf{x}_p$ approaches the boundary, $S$, can be now removed by subtracting the boundary regularised boundary integral equation (\ref{eq:BRIEF}) from (\ref{eq:CBIMpminuspsi}). Then using (\ref{eq:hlmhltz}) for $\psi(\mathbf{x})$ we obtain a numerically robust expression for $p(\mathbf{x}_p)$ that is free of near singular behaviour irrespective of the distance from $\mathbf{x}_p$ to any boundary:
\begin{align} \label{eq:BRIEFdmn}
4&\pi p(\mathbf{x}_p) = 4\pi \left[  p(\mathbf{x}_0)g({\mathbf{x}_{p}})  +  \left(\frac{\partial p}{\partial n}\right)_{0} f({\mathbf{x}_{p}}) \right] \nonumber\\
&-\int_{S + S_{\infty}} \left[p(\mathbf{x}) - p(\mathbf{x}_0)g({\mathbf{x}})  -  \left(\frac{\partial p}{\partial n}\right)_{0} f({\mathbf{x}}) \right] \left[\frac{\partial G(\mathbf{x}_p, \mathbf{x})}{\partial n}-\frac{\partial G(\mathbf{x}_0, \mathbf{x})}{\partial n}\right] \text{d}S(\mathbf{x})  \nonumber \\
&+ \int_{S + S_{\infty}} \left[ \frac{\partial p(\mathbf{x})}{\partial n} - p(\mathbf{x}_0) \nabla g(\mathbf{x}) \cdot \mathbf{n}(\mathbf{x}) - \left(\frac{\partial p}{\partial n}\right)_{0} \nabla f(\mathbf{x}) \cdot \mathbf{n}(\mathbf{x}) \right] \left[G(\mathbf{x}_p, \mathbf{x})-G(\mathbf{x}_0, \mathbf{x}) \right] \text{d}S(\mathbf{x}).
\end{align}
%Analytic expressions for the integrals over $S_{\infty}$ are given in Appendix \ref{sec:appA}.
%
% ======================================================

\section{Examples from acoustics}
We use examples drawn from acoustics to illustrate the implementation and advantages of the BRIEF of the Helmholtz equation relative to the conventional boundary integral method (CBIM). The first classic problem is the pressure field outside a radiating sphere of radius, $R$, with a prescribed time harmonic, radially symmetric surface normal velocity $v_n = V \exp(-i \omega t)$ \cite{Schenck1968}, giving the boundary condition: $\partial p / \partial n =- i \omega \rho V$, with $k = \omega / c$ where $\rho$ is the density and $c$ the speed of sound in the region outside the sphere. This is a Neumann problem since $p(\mathbf{x})$ is to be found for prescribed values of the normal derivative $\partial p / \partial n$. The analytic solution of this problem is \cite{Schenck1968}: $p(r)=-[(i \omega\rho V R^2)/(1-ikR)] e^{ik(r-R)}/r$.

  % -- Figure 1
\begin{figure}[t]
\centering
\includegraphics[width=1.0\textwidth] {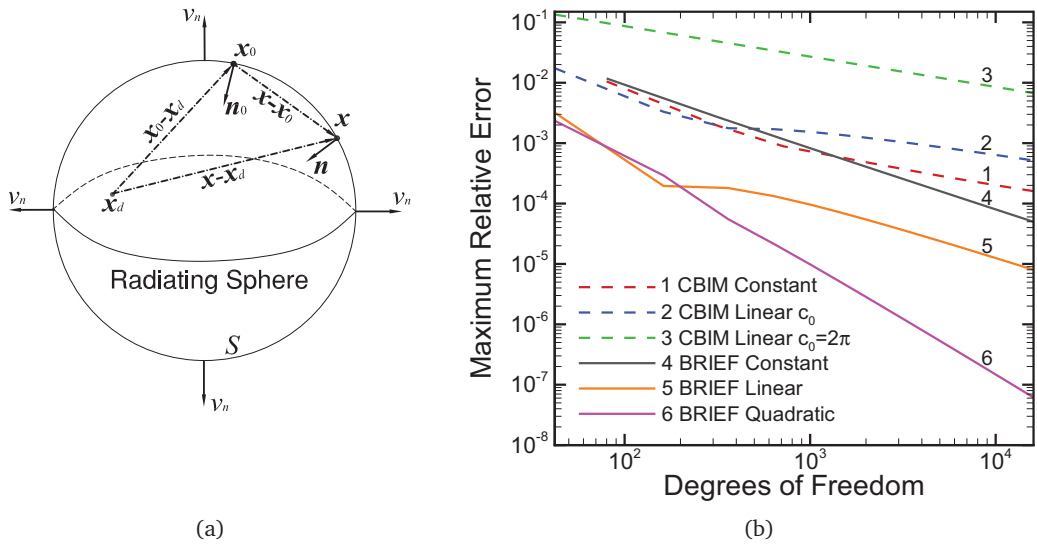}
\caption{(a) The radiating sphere, $S$ with $\mathbf{x}_0$, $\mathbf{x}$; surface normals $\mathbf{n}_{0}$ and $\mathbf{n}$; the point $\mathbf{x}_d$ lies inside $S$; (b) Maximum relative error of six numerical approaches as functions of the number of surface nodes or degrees of freedom, for $kR=\pi/2$ and $\mathbf{x}_d$ in (\ref{eq:gANDf}) is at the origin of the sphere.}
\label{fig:logerr}
\end{figure}

In figure 1b, we compare the maximum relative error of the following six different numerical solutions of the above problem at $kR = \pi/2$ for varying numbers of degrees of freedom or unknowns on the surface of the sphere:
\renewcommand{\theenumi}{\arabic{enumi}.}
\begin{enumerate}
\item \textsf{CBIM Constant}: The value of the unknown pressure associated with every \emph{flat} triangular surface element is assumed to be \emph{constant} within the element.
\item \textsf{CBIM Linear} $c_0$: The points on the surface of the sphere on which the pressure is to be found are at the vertices of \emph{planar} triangular area elements with linear shape functions used to represent the boundary. The solid angle, $c_0$ in (\ref{eq:cnbimhlmhltz}) is calculated according to the local geometry around each point $\mathbf{x}_0$ \cite{Mantic93}.
\item \textsf{CBIM Linear} $c_0 = 2\pi$: This an often used approach that assumes the \emph{incorrect} value of $c_0=2\pi$ for the solid angle at every node even though the surface tangent plane is clearly not defined at the vertices of the surface elements.
\item \textsf{BRIEF Constant}: Using BRIEF with the assumption that the unknown pressure associated with every \emph{flat} triangular surface element is \emph{constant} within the element.
\item \textsf{BRIEF Linear}: Using BRIEF where the points on the surface of the sphere on which the pressure is to be found are at the vertices of \emph{planar} triangular area elements with linear shape functions used to represent the boundary.
\item \textsf{BRIEF Quadratic}: Using BRIEF where the points on the surface of the sphere on which the pressure is to be found are at the nodes of \emph{quadratic} 6-noded triangular area elements with quadratic shape functions used to represent the boundary.
\end{enumerate}

It is clear from figure 1b that results obtained from the BRIEF are at least as good as, and in the case of \textsf{BRIEF Quadratic}, far superior to any results from the CBIM. Depending on the degrees of freedom, the relative error is smaller by 1 to 3 orders of magnitude; or \textsf{BRIEF Quadratic} can achieve the same precision as any CBIM with about $1/10th$ the number of degrees of freedom. In all BRIEF approaches, there is no need to deal with singular integrals or to compute the solid angle $c_0$ at any node. Moreover, the poor performance of the erroneous \textsf{CBIM Linear} $c_0 = 2\pi$ approach is demonstrated.

  % -- Figure 2
\begin{figure}[t]
\centering
\includegraphics[width=1.0\textwidth] {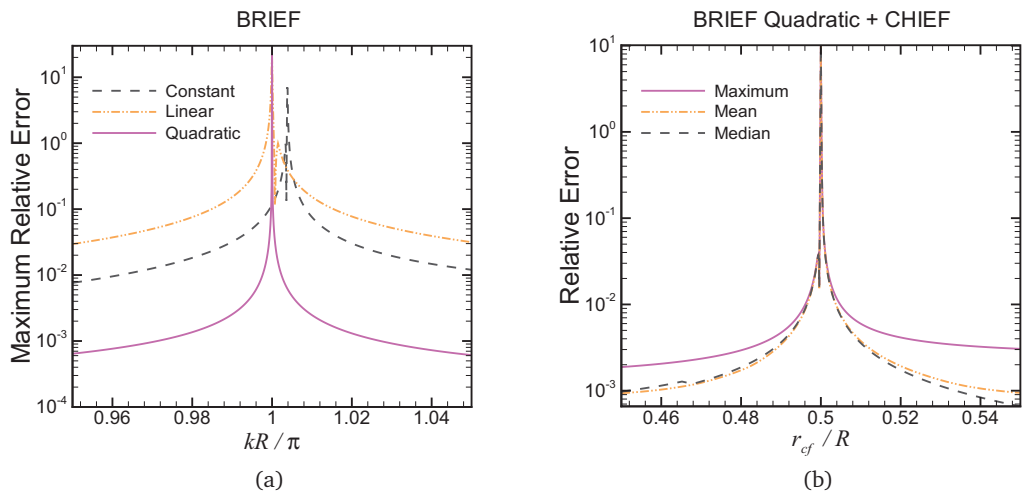}
\caption{Solutions of the single radiating sphere problem in figure 1 using the BRIEF with 1280 constant elements (dash line) and 624 nodes with 1280 linear (dash dot line) or 320 quadratic (solid line) elements. (a) Maximum relative error as a function of the wave number, $k$; (b) Maximum, mean and median relative errors of results from \textsf{BRIEF Quadratic} plus CHIEF as a function of the location of the CHIEF point, $r_{cf}$ for $kR=2\pi$.  The point $\mathbf{x}_d$ in (\ref{eq:gANDf}) is at the origin of the sphere.}\label{fig:Maxerrwn}
\end{figure}

In figure 2a, we compare the BRIEF using constant, linear and quadratic elements for a single radiating sphere considered at wave numbers very close to the first resonant value, $kR=\pi$ \cite{Schenck1968}. We observe that: (i) using the BRIEF with quadratic (\textsf{BRIEF Quadratic}) elements, the maximum relative error is not significant until $|kR-\pi| < 10^{-4}$; (ii) when the BRIEF is used with constant elements (\textsf{BRIEF Constant}), the resonant wave number is located incorrectly. This is because the spherical shape is not well represented by constant elements and resulted in a slightly different resonant wave number, even though the polyhedra that represent the sphere have the same volume. Even in this case, the resonant solution is only evident when $k$ is within 1\% of the (incorrect) resonant value.

In Table 1, we compare the condition number at various values of $kR$. We see that using \textsf{BRIEF Linear} or \textsf{BRIEF Quadratic}, the condition number at $kR$ very close to the resonant value can be reduced by a factor of 60 when compared to \textsf{BRIEF Constant}. This demonstrates the numerical stability that can be achieved using \textsf{BRIEF Linear} or \textsf{BRIEF Quadratic}.

 % -- TABLE 1 --
\begin{table*}[bp]\centering
\caption{Variation of the condition number of the BRIEF for a pulsating sphere of radius $R$ for two values of $kR$ close to resonance at $kR = \pi$, using constant, linear or quadratic surface elements to represent the sphere. The constant element representation predicts an incorrect resonant value very close to $kR/\pi = 1.0038$ - see also Figure 2.}\label{Tbl:1}
    \ra{1.3}
    \begin{tabular}{c c c}
        \hline\hline
         & $kR/\pi$ = 0.95 & $kR/\pi$ = 1.0038 \\
        \hline\hline \noalign{\smallskip}\noalign{\smallskip}
        Constant element  & 6.0 & 6000  \\
        \noalign{\smallskip}\noalign{\smallskip}
        \hline \noalign{\smallskip}\noalign{\smallskip}
        Linear element  & 6.5 & 81  \\
        \noalign{\smallskip}\noalign{\smallskip}
        \hline \noalign{\smallskip}\noalign{\smallskip}
        Quadratic element  & 6.5 & 100  \\ \noalign{\smallskip}\noalign{\smallskip}
        \hline\hline
    \end{tabular}
\end{table*}

Next, we study the error when \textsf{BRIEF Quadratic} is combined with one CHIEF point located at $r_{cf}$ inside the sphere. With $kR = 2\pi$ the lowest resonant mode has a node at $r = R/2$ \cite{Schenck1968}. The maximum, mean and median errors when the CHIEF point, $r_{cf}$ is positioned close to this node at $R/2$ are compared in figure 2b. Note that to incur a maximum error exceeding 1\%, $r_{cf}/R$ has to be within 0.004 of the node at $r/R =1/2$. Thus in practice, the \textsf{BRIEF Quadratic} approach plus one CHIEF point is very unlikely to encounter problems with the resonant solution.

  % -- Figure 3
\begin{figure}[t]
\centering
\includegraphics[width=1.0\textwidth] {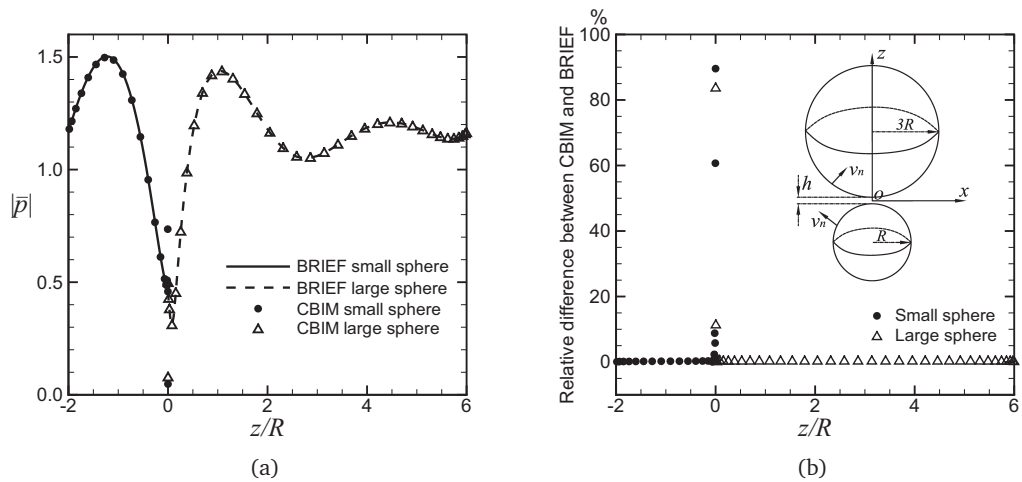}
\caption{Two spheres of radius $R$ and $3R$ at separation $h/R= 0.001$ radiating with a phase difference of $\pi$ calculated using $kR=\pi/2$ and 6938 nodes on each sphere. (a) The magnitude of the scaled pressure, $|\bar{p}| = |[(1-ikR)/(\rho \omega R V)] p|$ on the sphere surfaces along the longitudes through the point of closest approach. (b) Relative difference in the pressure magnitude, $|p|$ on the spheres obtained using \textsf{BRIEF Linear} and \textsf{CBIM Linear} along the longitudes through the point of closest approach. Inset: The geometry of the two spheres. The point $\mathbf{x}_d$ in (\ref{eq:gANDf}) is at the origin of each of the spheres.}\label{fig:2spheres}
\end{figure}

In a second example, we illustrate the unique ability of the BRIEF in being able to handle boundaries with extreme geometric aspect ratios by considering two radiating spheres that are nearly touching. We calculate the pressure field due to these spheres of radii $R$ and $3R$ at a distance of closest approach, $h/R = 0.001$, with $kR=\pi/2$ (see figure 3). The spheres have identical time harmonic radially symmetric normal velocities as in the previous example, but are out of phase by $\pi$. In figure 3a, we show the variation of the magnitude of the pressure, $|p|$, along the longitudes of the 2 spheres that pass through the point of closest approach at $z = 0$. Near this point, the result from \textsf{BRIEF Linear} remains continuous as expected whereas the result from \textsf{CBIM Linear} exhibits large errors that have their origin from the adverse influence of the nearly singular behaviour of the kernel of one surface on an adjacent surface that is very close by. The relative difference in the pressure magnitude, $|p|$ obtained using \textsf{BRIEF Linear} and \textsf{CBIM Linear} is shown in figure 3b.

A comparison of the accuracy in the the magnitude of the pressure obtained using the boundary regularised integral equation formulation (BRIEF) and the conventional boundary integral method (CBIM) is given in the contour plots for two spheres of radius $R$ and $3R$, at separation $h/R = 0.001$ as illustrated in figure 3b. In figure 4a, we compare the variation of the magnitude of the pressure in the median plane, $z=0$, in the neighbourhood of the osculating point and in figure 4b, pressure variations in the far field. Around $\rho \equiv (x^2 + y^2)^{1/2} \sim 0$, the {\textsf{BRIEF Linear}} results remain smooth as expected from the results in figure 3a. In contrast, the numerical errors of the conventional boundary integral method (CBIM) are already quite evident from the roughness of the pressure contours at $\rho \sim 0.7R$, and these errors become unacceptably large for $\rho < R/2$. This is consistent with the comparison shown in figure 3. The pressure variations in the far field for the two-spheres in the plane $y=0$ shown in figure 4b demonstrate the utility of the BRIEF in furnishing accurate results in both the near and far field.

  % -- Figure 4
\begin{figure}[t]
\centering
\includegraphics[width=1.0\textwidth] {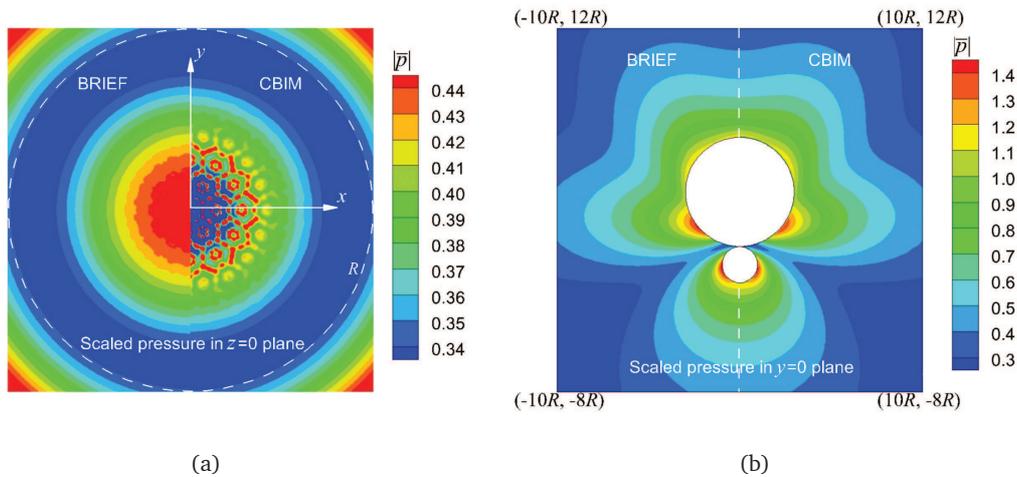}
\caption{Contour plots of the scaled pressure magnitude, $|\bar{p}| = |[(1-ikR)/(\rho \omega R V)] p|$,  obtained by the boundary regularised integral equation formulation (\textsf{BRIEF Linear}) and by the conventional boundary integral method (\textsf{CBIM Linear}) for the two closely spaced radiating spheres of radius $R$ and $3R$, at separation $h/R = 0.001$: (a) pressure contours in the median plane, $z=0$, within square region $|x|<R$, $|y|<R$, and (b) pressure contours in the far field in the $y=0$ plane. Linear elements and 6938 nodes were used on each sphere in both cases. All other parameters are the same as those in figure 3.}
\label{fig:2spheresz0}
\end{figure}

%================================================================================================
\section{Conclusions}
We have developed the boundary regularised integral equation formulation (BRIEF) of the solution of the Helmholtz equation given by (\ref{eq:BRIEF}) that replaces the conventional boundary integral method (CBIM) given by (\ref{eq:cnbimhlmhltz}). As all integrands in the BRIEF contain no singular behaviour and the term containing the solid angle has been removed, any convenient quadrature method can be used to evaluate the surface integrals thus facilitating the use of more accurate linear, quadratic or spline elements to represent the boundaries.

The BRIEF also has the unique ability to handle problems with extreme geometric aspect ratios such as where boundaries are very close together or where boundaries possess very different characteristic length scales. For such cases, the absence of singular behaviour in the BRIEF means that high precision is always maintained whereas the inherent singularities in the integral of the CBIM will give rise to unavoidable deteriorations in numerical precision.

The absence of singular behaviour in the surface integrals means that the BRIEF provides a simple and numerically robust way to compute the solution at field points located near boundaries using (\ref{eq:BRIEFdmn}). This is an important advance in the accurate evaluation of the solution near surfaces or within small gaps between surfaces where the errors in the conventional boundary integral method become unacceptably large using the same surface mesh. This is an advantage of the BRIEF that is unmatched by the CBIM without very significant additional numerical and analytical effort.

In the derivation of the key result, (\ref{eq:BRIEF}), we have assumed that the boundary $S$ is sufficient smooth whereby the tangent plane is defined at all points on the surface. For points on boundaries at which the tangent plane is not uniquely defined, for example at sharp corners and edges, one can still implement BRIEF using the double node technique to construct the requisite system of linear equation \cite{Sun2014}.

From numerical case studies drawn from acoustics considered here, the use of linear elements or quadratic elements with the BRIEF can lower the relative error by a factor of 10 to 1000. Conversely, the same precision as the CBIM can be obtained even when the number of degrees of freedom is reduced by a factor of 10 to 100 or more. This represents a very significant speed up in computational time.

Of particular relevance to acoustic problems, the much higher precision results that can be obtained using the BRIEF also means that the uniqueness issue associated with the resonant or normal mode solutions of the Helmholtz equation around closed surfaces is very unlikely to arise. This is simply because the numerical value of the wave number must be within a relative deviation of smaller than $10^{-3}$ from the resonant value before the resonant solution can start to contribute to the answer. Indeed, if the CHIEF method of Schenck \cite{Schenck1968} is also used, the uniqueness issue will not arise in practice since the CHIEF node must now lie within a relative deviation of smaller than $10^{-3}$ of an internal resonance node before the BRIEF-CHIEF combination might breakdown. The BRIEF is also much simpler than the Burton-Miller \cite{Burton1971} approach because no hypersingular integrals are involved.

Given the above advantages of the BRIEF, the case for its wide adoption is very compelling, not only for the Helmholtz equation that has applications ranging from wave phenomena in acoustics to electromagnetics, but also for boundary integral solutions of the Laplace equation \cite{Sun2014} and hydrodynamic problems \cite{Sun2013}.

%================================================================================================
\section*{Acknowledgment}
This work is supported in part by the Australian Research Council Discovery Project Grant Scheme to DYCC who is a Visiting Scientist at the Institute of High Performance Computing and an Adjunct Professor at the National University of Singapore.

%================================================================================================
\appendix
\section{Derivation of the integrals over the surface at infinity}\label{sec:appA}
In this Appendix, we derive analytic expressions for the integrals over the surface at infinity, $S_{\infty}$, that appear in (\ref{eq:intinf1}), (\ref{eq:intinf2}) and (\ref{eq:BRIEFdmn}).

In (\ref{eq:intinf1}) and (\ref{eq:intinf2}) the integrals at infinity vanish due to the Sommerfeld radiation condition \cite{Sommerfeld12, Schot92}:
\begin{align}\label{eq:sommerfeld}
    r \left( \frac{\partial p} {\partial r} - i k p\right) \rightarrow 0,  \quad \text{as}  \quad r \rightarrow \infty,
\end{align}
except for the two terms
\begin{align}\label{eq:ginf}
\int_{S_{\infty}} \left(\frac{\partial{g}}{\partial{n}}G-g\frac{\partial{G}}{\partial{n}}\right)\text{ d}S
\end{align}
and
\begin{align}\label{eq:finf}
\int_{S_{\infty}} \left(\frac{\partial{f}}{\partial{n}}G-f\frac{\partial{G}}{\partial{n}}\right)\text{ d}S
\end{align}
that correspond to Eqs. (\ref{eq:intinf1}) and (\ref{eq:intinf2}), respectively. For simplicity, we have applied $g\equiv g(\mathbf{x})$, $f\equiv f(\mathbf{x})$, $\partial{g}/\partial{n}\equiv\nabla g(\mathbf{x}) \cdot \mathbf{n}(\mathbf{x})$, $\partial{f}/\partial{n}\equiv\nabla f(\mathbf{x}) \cdot \mathbf{n}(\mathbf{x})$, $G\equiv G(\mathbf{x}_0,\mathbf{x})$, $\partial{G}/\partial{n}\equiv \partial{G(\mathbf{x}_0,\mathbf{x})}/\partial{n}$ and $\text{ d}S\equiv\text{ d}S(\mathbf{x})$.

Both $g$ and $f$ vanish as $1/r_{d} \sim 1/r$ at the surface at infinity, $S_{\infty}$, and $G$ and $\partial{G}/\partial{n}$ vanish as $1/r$, when $r\rightarrow \infty$, because
\begin{align}
G=\frac{e^{ikr}}{r},
\end{align}
and
\begin{align}
\lim_{r\rightarrow\infty}\frac{\partial{G}}{\partial{n}} = \lim_{r\rightarrow\infty}\frac{e^{ikr}}{r^3}(ikr-1)\mathbf{n} \cdot \mathbf{x} \rightarrow ik\frac{e^{ikr}}{r},
\end{align}
where terms that vanish faster than $1/r$ can be omitted because the surface $S_{\infty}$ only grows as $r^{2}$ at infinity. Consequently, only terms multiplying $\partial{G}/\partial{n}$, that vanish as $1/r$ can give contributions to the integrals.

The sine and cosine terms in $g$ and $f$, see (\ref{eq:gANDf}), can be rewritten in complex notation as
\begin{align}\label{eq:gcmplx}
g=\frac{e^{ik(r_{d}-a)}}{2r_{d}}\left[a-\frac{i}{k}\right] +\frac{e^{-ik(r_{d}-a)}}{2r_{d}}\left[a+\frac{i}{k}\right],
\end{align}
\begin{align}\label{eq:fcmplx}
f=\frac{-ia}{2bkr_{d}}\left[e^{ik(r_{d}-a)}-e^{-ik(r_{d}-a)}\right].
\end{align}
Therefore the normal derivatives of $g$ and $f$ as $r \rightarrow \infty$ asymptotes to :
\begin{align}
\lim_{r\rightarrow\infty} \frac{\partial{g}}{\partial{n}} \rightarrow \frac{ik}{2r_{d}}\left[a-\frac{i}{k}\right]e^{ik(r_{d}-a)} - \frac{ik}{2r_{d}}\left[a+\frac{i}{k}\right]e^{-ik(r_{d}-a)},
\end{align}
\begin{align}
\lim_{r\rightarrow\infty} \frac{\partial{f}}{\partial{n}} \rightarrow \frac{a}{2br_{d}}\left[e^{ik(r_{d}-a)}+e^{-ik(r_{d}-a)}\right]
\end{align}
where we have again neglected higher order terms.

It can be easily seen that the terms proportional to $e^{ik(r_{d}-a)}$ in the integrals of (\ref{eq:ginf}) and (\ref{eq:finf}) remain finite and the individual terms cancel out exactly. However, for the terms proportional to $e^{-ik(r_{d}-a)}$, this is not the case and they contribute to the integrals in (\ref{eq:ginf}) and (\ref{eq:finf}) to give:
\begin{align}
\int_{S_{\infty}} \left(\frac{\partial{g}}{\partial{n}}G-g\frac{\partial{G}}{\partial{n}}\right)\text{ d}S = \frac{a}{b}\int_{S_{\infty}} \frac{e^{ik(r-r_{d}+a)}}{rr_{d}}\text{ d}S,
\end{align}
and
\begin{align}
\int_{S_{\infty}} \left(\frac{\partial{f}}{\partial{n}}G-f\frac{\partial{G}}{\partial{n}}\right)\text{ d}S = -ika\left[1+\frac{i}{ka}\right]\int_{S_{\infty}}\frac{e^{ik(r-r_{d}+a)}}{rr_{d}}\text{ d}S.
\end{align}
Finally we need to determine the integral
$$\int_{S_{\infty}} \frac{e^{ik(r-r_{d}+a)}}{rr_{d}}\text{ d}S$$
that appears in both of the above equations. The denominator $rr_{d}$ of the integrand can be approximated by $r^2$ since $r\approx r_{d}$. For the numerator, we need to invoke the cosine rule and a Taylor expansion to $r-r_{d}+a$, which gives $r-r_{d}+a \approx a(1+\cos{\alpha})$ with $\alpha$ being the angle between the line segments $r_{d}$ and $a$. We then get:
\begin{align}\label{eq:intinfkeyG}
\int_{S_{\infty}} \frac{e^{ik(r-r_{d}+a)}}{rr_{d}}\text{ d}S = \int_{S_{\infty}} \frac{e^{ika(1+\cos{\alpha})}}{r^2}\text{ d}S = \int_{0}^{\pi}\frac{e^{ika(1+\cos{\alpha})}}{r^2}2\pi r^2 \sin{\alpha} \text{ d}\alpha = -\frac{2\pi}{ika}\left[1-e^{2ika}\right].
\end{align}
This leads immediately to (\ref{eq:intinf1}) and (\ref{eq:intinf2}).

In (\ref{eq:BRIEFdmn}), we need to evaluate the following two integrals over the surface at infinity, $S_{\infty}$:
\begin{align}\label{eq:gpinf}
\int_{S_{\infty}} \left(\frac{\partial{g}}{\partial{n}}G^{p}-g\frac{\partial{G^{p}}}{\partial{n}}\right)\text{ d}S,
\end{align}
and
\begin{align}\label{eq:fpinf}
\int_{S_{\infty}} \left(\frac{\partial{f}}{\partial{n}}G^{p}-f\frac{\partial{G^{p}}}{\partial{n}}\right)\text{ d}S
\end{align}
where $G^{p}\equiv G(\mathbf{x}_p,\mathbf{x})$, $\partial{G^{p}}/\partial{n}\equiv \partial{G(\mathbf{x}_p,\mathbf{x})}/\partial{n}$. Here we can rewrite $g$ and $f$ in (\ref{eq:gcmplx}) and (\ref{eq:fcmplx}) as
\begin{align}\label{eq:gpcmplx}
g=\frac{e^{ik[(r_{d}-a_{p})+(a_{p}-a)]}}{2r_{d}}\left[a-\frac{i}{k}\right] +\frac{e^{-ik[(r_{d}-a_{p})+(a_{p}-a)]}}{2r_{d}}\left[a+\frac{i}{k}\right],
\end{align}
\begin{align}\label{eq:fpcmplx}
f=\frac{-ia}{2bkr_{d}}\left[e^{ik[(r_{d}-a_{p})+(a_{p}-a)]}-e^{-ik[(r_{d}-a_{p})+(a_{p}-a)]}\right],
\end{align}
in which $a_{p}=|\mathbf{x}_p-\mathbf{x}_d|$. Following the same procedure we used to evaluate (\ref{eq:ginf}) and (\ref{eq:finf}), the key integral that needs to be determined for calculating of (\ref{eq:gpinf}) and (\ref{eq:fpinf}) is
$$\int_{S_{\infty}} \frac{e^{ik(r_{p}-r_{d}+a_{p})}}{r_{p}r_{d}}\text{ d}S,$$
where $r_{p} = |\mathbf{x}-\mathbf{x}_p|$. By using (\ref{eq:intinfkeyG}), we then find
\begin{align}\label{eq:intinfkeyGp}
\int_{S_{\infty}} \frac{e^{ik(r_{p}-r_{d}+a_{p})}}{r_{p}r_{d}}\text{ d}S  = -\frac{2\pi}{ika_{p}}\left[1-e^{2ika_{p}}\right].
\end{align}

%%%%%%%%%% Insert bibliography here %%%%%%%%%%%%%%

\end{document}